\documentclass[aps,prl,twocolumn,groupedaddress]{revtex4}

\usepackage{graphicx}
\usepackage{dcolumn}
\usepackage{bm}

\begin{document}
\title{Experimental search for anisotropic flux flow resistivity\\
in the \textit{a-b} plane of optimally doped epitaxial thin films
of $YBa_2Cu_3O_{7-\delta}$}

\author{G. Koren, P. Aronov and E. Polturak}
\affiliation{Department of Physics, Technion - Israel Institute of
Technology, Haifa 32000, Israel}

\email{gkoren@physics.technion.ac.il}
\homepage{http://physics.technion.ac.il~gkoren}

\date{\today}

\begin{abstract}

Transport measurements along the node and anti-node directions in
the {\it a-b} plane of optimally doped and epitaxial thin films of
$YBa_{2}Cu_{3}O_{7-\delta}$ are reported. Low bias
magnetoresistance measurements near and below $T_c$ show that the
flux flow resistivity along the node and anti-node directions
versus magnetic field are indistinguishable. This result suggests
that within the experimental error of our measurements,
\textit{no} correspondence is found between the flux pinning
properties in YBCO and the d-wave nature of the order parameter.

\end{abstract}

\pacs{74.25.Qt, 74.72.Bk, 74.50.+r }

\maketitle

\indent Under applied magnetic field, the apparent resistance of
type II superconductors in the mixed state is due to the motion of
vortices. In the presence of transport current, the Lorentz (or
Magnus) force on vortices causes motion and induces a voltage drop
across the superconducting sample. The ratio of the induced
voltage divided by the current defines the flux flow resistance
\cite{Huebener,TinkhamA}. In the Bardeen-Stephen model
\cite{Bardeen-Stephen}, valid for the case of weak pinning as for
instance near $T_c$, the effective viscosity associated with the
motion of vortices is caused by the interaction of the transport
current and the shielding currents around the vortex. In the
cuprates, the shielding currents around the vortex are predicted
to show anisotropy linked to the d-wave nature of the order
parameter \cite{Ichioka,Mizel}. Under the quasiclassical
approximation and by the use of the Eilenberger equations, Ichioka
\textit{et al} found a small four fold anisotropy of a few
percents in the induced supercurrents around the vortex core
\cite{Ichioka}. A preliminary microscopic calculation of vortex
tunneling also seems to yield an \textit{anisotropic} vortex
dynamics \cite{Mizel}. It is therefore plausible to assume that
the effective viscosity which determines the flux flow resistivity
may also be anisotropic. The search for this effect is the subject
of the present study. We designed a specific, high precision
experiment to look for it by Magnetotranport measurements which
were conducted on nominally identical thin film microbridges of
YBCO patterned on the same wafer along the node and antinode
directions of the d-wave order parameter. We found that the flux
flow resistance at low bias did not reveal any anisotropy at the
1\% level, which is the stated precision of our measurements.\\

\indent To facilitate the comparison between the transport
properties along the node and antinode directions, two high
quality \textit{c-axis} oriented epitaxial thin films of YBCO were
prepared under identical conditions by laser ablation deposition
on (100) $SrTiO_3$ (STO) wafers of $10\times 10\,mm^2$ area. In
one case, the orientation of the single crystal STO substrate was
with the edge of the wafer parallel to the (010) crystalline
direction, while in the other case it was parallel to the (110)
direction. Ten microbridges were defined on each wafer using the
same photolithographic mask, and patterned by Ar ion milling at a
temperature of  -170 $^\circ$C. The dimensions of the microbridges
were $0.12\times 12\times 100\, \mu m^3$. Successive microbridges
were oriented at alternating angles of 0$^\circ$ and 45$^\circ$ to
the edge of the wafers, so that the transport current would flow
either along the node or the antinode of the order parameter. The
alternating direction of \textit{adjacent} microbridges was
important to minimize systematic differences due to possible
inhomogeneities in the films.  On the first wafer with the side
parallel to the (010) orientation, five odd number bridges were
along the antinode direction, and five even number bridges along
the node direction. In the second wafer with the side parallel to
the (110) orientation, the role of the antinode and node bridges
was reversed due to the epitaxial growth of the film.  Studying
these two types of wafers was done in order to check if our ion
milling process, done at an incident angle of 45$^\circ$ to the
wafers, is affecting the properties of the microbridges. We note
that on the first wafer, the Ar ions milling process leads to
antinode bridges with sides normal to the surface, but at an
oblique angle for the node microbridges. This situation is
reversed in the second wafer. Any observed difference in the
transport properties of the two wafers would imply that the effect
is not intrinsic, and results from the patterning process. Low
resistance gold contacts were prepared on the two wafers by laser
ablation deposition and lift off, followed by further oxygen
annealing for the gold/YBCO contact (at 650$^\circ$C) and the YBCO
films themselves (at 450$^\circ$C). Transport measurements were
done by the standard 4-probe dc technique, with and without a
magnetic field of up to 8T normal to the wafers (parallel to the
\textit{c-axis}).\\
\begin{figure}
\includegraphics[height=6.5cm,width=9cm]{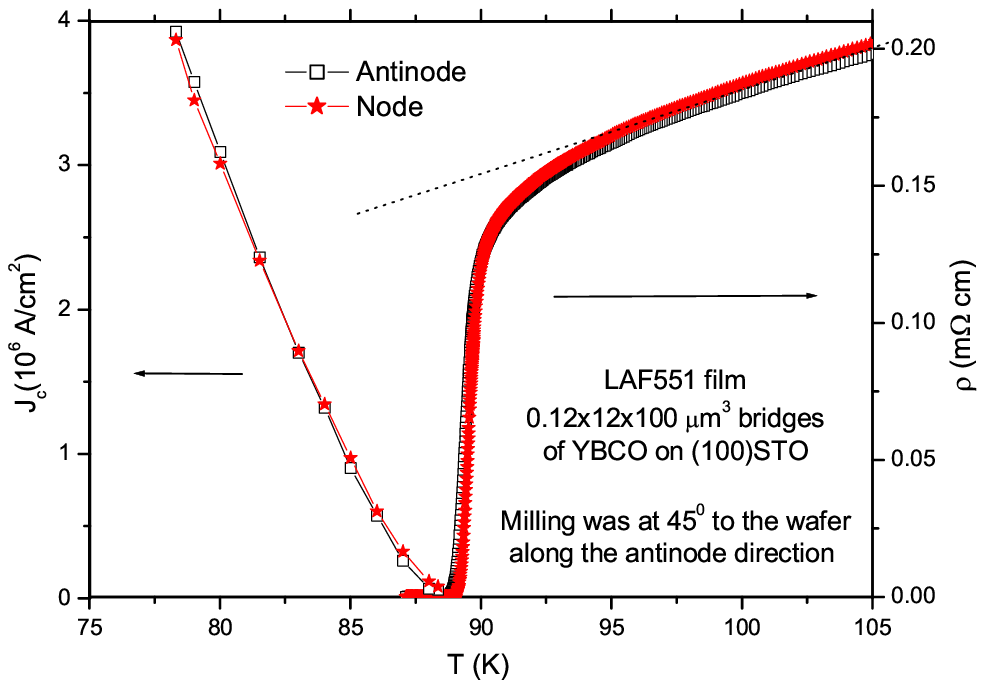}
\caption{\label{fig:epsart} (Color on-line) Mean resistivity and
critical current density versus temperature of three microbridges
of YBCO along an anti-node direction (squares), and three along a
node direction (stars). The antinode bridges are parallel to the
(100) side of the STO wafer, and the node bridges are oriented at
45$^\circ$ to it. The lines connecting the data points are guide
to the eye. The straight dashed line extrapolating the normal
state resistivity to lower temperatures shows that $T_c$(onset) is
$\sim 94K$. }
\end{figure}
\begin{figure}
\includegraphics[height=6.5cm,width=9cm]{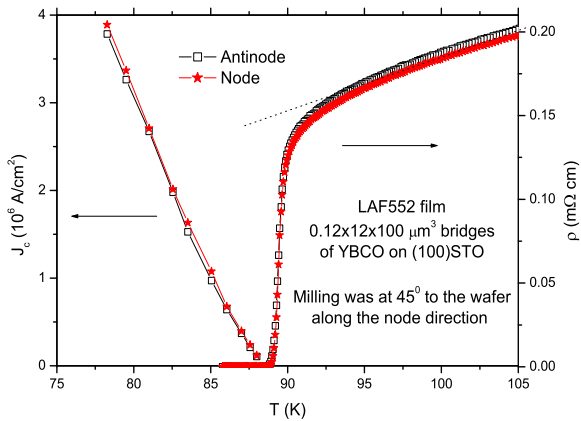}
\caption{\label{fig:epsart} (Color on-line) Mean resistivity and
critical current density versus temperature of three microbridges
of YBCO along an anti-node direction (squares), and three along a
node direction (stars). The node bridges are parallel to the (110)
side of the STO wafer, and the node bridges are oriented at
45$^\circ$ to it. The lines connecting the data points are guide
to the eye.}
\end{figure}

\indent Figures 1 and 2 show mean values of the zero field normal
state resistivity $\rho$ and critical current density $J_c$ on the
two wafers, as a function of temperature. The mean values were
obtained for each wafer by averaging over three microbridges of
each type (node or antinode) which had the closest $T_c$(R=0)
values. The critical current data was measured using a $1\,\mu$V
per $100\,\mu$m bridge length criterion.  The normal state
resistivities along the node and antinode directions are slightly
different, by about 2\%. In Fig. 1 the node resistivity is higher,
and in Fig. 2 the opposite situation is found where the antinode
resistivity is higher. This behavior results from the fact that
the ion milling process slightly damages the side of the bridges
which are exposed to the Ar ion beam, the node bridges in Fig. 1
and the antinode bridges of Fig. 2. Apart from this minor
difference, $\rho$(node) should be equal to $\rho$(anti-node)
since our films are heavily twinned. This results from the fact
that due to twinning one has $\rho$=($\rho_a$+$\rho_b$)/2 for the
anti-node bridges, and $\rho=\rho_a cos^2(45^\circ)+\rho_b
sin^2(45^\circ)$ for the node bridges, which are of course equal.
The transition temperatures $T_c$(onset) in Figs. 1 and 2 are
identical for both type of bridges. $T_c$(onset)$\sim 94$K is the
temperature at which the resistivity data in Figs. 1 and 2
deviates from the straight dashed line extrapolating the high
temperature data to lower temperatures. The transition
temperatures of zero resistance $T_c$(R=0) of the node and
antinode bridges are very close. In Fig. 1 the values are 88.9K
for the antinode bridges and 89.1K for the node ones, while in
Fig. 2 the corresponding values are 88.8K and 88.9K.  Although
within the experimental noise, the slightly higher $T_c$ by about
0.1\% of the node microbridges as compared to the antinode ones,
is consistent with Ichioka \textit{et al.} results where the
supercurrent around a vortex is slightly higher along the node
directions \cite{Ichioka}. The differences between the node and
anti-node bridges are also very small in the critical current
results, where near $T_c$ the small difference in $T_c$ affects
the results, but at lower temperatures this difference is within
the experimental noise. For temperatures close to $T_c$(R=0),
$J_c$(T) increases with decreasing temperature as $(T_c-T)^{3/2}$,
and linearly at lower temperatures with a slope of $0.45\times
10^6\,A/cm^2$K. So we can conclude from Figs. 1 and 2 that the
node and antinode bridges have almost identical $T_c$
and $J_c(T)$.\\
\begin{figure}
\includegraphics[height=6.5cm,width=9cm]{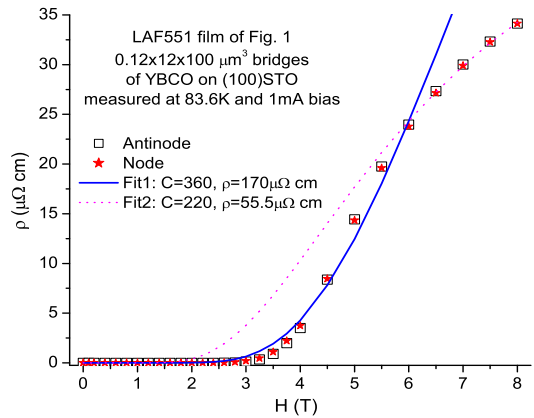}
\caption{\label{fig:epsart} (Color on-line) Mean flux flow
resistivity at 83.6K versus applied magnetic field of the bridges
of Fig. 1. The mean values are of three bridges along the node
direction (stars), and three along the anti-node direction
(squares). The two curves are fits using the Tinkham's model as
given by Eq. (1).}
\end{figure}
\begin{figure}
\includegraphics[height=6.5cm,width=9cm]{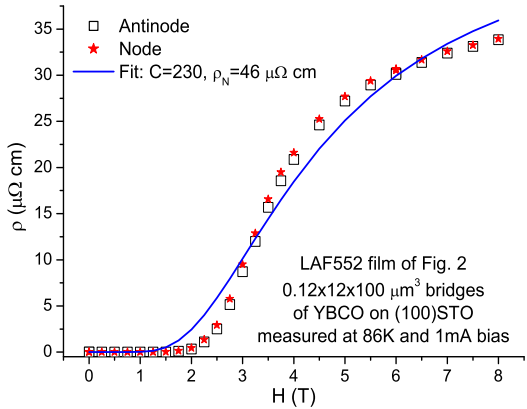}
\caption{\label{fig:epsart} (Color on-line) Mean flux flow
resistivity at 86K versus applied magnetic field of the bridges of
Fig. 2. The mean values are of two bridges along the node
direction (stars), and two along the anti-node direction
(squares). Mean values were taken here over two and not three
microbridges due to contact problems with one of the bridges. The
curve is a fit using the Tinkham's model as given by Eq. (1).}
\end{figure}

\indent Fig. 3 and Fig. 4 show the flux flow resistivity of
microbridges of Fig. 1 and Fig. 2, respectively, as a function of
applied magnetic field normal to the wafers.  The onset field at
which the flux flow resistivity first appears increases with
decreasing temperature, due to the stronger pinning of vortices at
lower temperatures. Figs. 3 and 4 show that the flux flow
resistivity curves versus field of both kinds of microbridges,
those along the node and anti-node directions are almost
indistinguishable.  This is the main experimental observation of
the present study. Both figures also show that at high fields the
flux flow resistivity tends to saturate, more so in Fig. 4 than in
Fig. 3 due to the higher temperature. This is possibly due to a
cross over to the normal state resistivity, although the measured
value just above the transition $\rho_N(95K)=170\,\mu\Omega cm$
(see Figs. 1 and 2) is much higher than the $40-50\,\mu\Omega cm$
value which can be extrapolated from Figs. 3 and 4. The overall
behavior of $\rho$ versus \textit{H} is consistent with previous
experiments on YBCO \cite{Ossandon,Kunchur}. From the linear part
of the data with the highest slope, one can extract the critical
field $H_{c2}$ by using the Bardeen-Stephen model
\cite{TinkhamA,Bardeen-Stephen,Ossandon,Kunchur}. The curves in
Figs. 3 and 4 are two Ambegaokar-Halperin type fits using
Tinkham's model \cite{Ambegaokar,Tinkham} which yields:
\begin{equation}
\rho_{ff}=\rho_N[I_0(U_0/2k_BT)]^{-2}=\rho_N[I_0(C(1-t)^{3/2}/H)]^{-2}
\end{equation}
where $\rho_{ff}$ and $\rho_N$ are the flux flow and normal state
resistivities, $I_0$ is the modified Bessel function, $U_0$ is the
activation energy, $t=T/T_c$ is the reduced temperature, and $C$
is a constant. The basic physics behind this model is that the
motion of vortices between pinning sites is thermally activated
and involves phase slippage of $2\pi$ like in a single heavily
damped current driven Josephson junction. One sees that the fits
can not reproduce the main features of the data all at once. In
Fig. 3, we chose to show that they can either fit the data
reasonably well up to a field of 6T but miss the higher fields
data (solid curve), or fit the higher fields but miss the onset
and the intermediate fields data (dashed curve). In Fig. 4 we
chose to show a single fit for the whole range of fields, but than
the fit quality is quite poor. We note that the low fields fit in
Fig. 3 (Fit1) is obtained with the actually measured normal state
resistivity just above $T_c$(onset) at 95K ($170\,\mu\Omega cm$).
We shall discuss the reliability and suitability of the Tinkham
model later on, here we point out that the important thing is that
the model allows us to obtain the activation energy $U_0$ which is
related to the fit parameter $C$ by $U_0=2k_BT_ct(1-t)^{3/2}C/H$
as seen in Eq. (1). Since the node and antinode curves in Figs. 3
and 4, respectively are almost identical, the same $U_0$ is
obtained for both orientations for each temperature. Therefore, at
any given temperature, the activation energy for moving vortices
along both directions is the same. This is the main result of the
present study. It indicates that the pinning properties of YBCO
which control the flux flow resistivity are fully isotropic to
within the experimental error, and are not affected by the
intrinsic anisotropy of the d-wave order parameter as might be
expected.\\

\indent Finally, we discuss the suitability of using the Tinkham
model given by Eq. (1) for the present results. We have already
seen the problems involved in using this model in the fits of
Figs. 3 and 4. To elucidate this issue, we show in Fig. 5 the flux
flow resistivity data versus magnetic field of a microbridge on
LAF552 with a weak link in it (a scratch). The weak link leads to
a very small critical current in this bridge, and the onset of the
flux flow resistivity occurs at a very small field. The solid
curve is a fit of this data using Eq. (1). One can see that in
this bridge the Tinkham model of vortex motion by thermal
activation fits the data quite nicely except for fields below
about 1.5T. In this regime a different mechanism must be involved
in the resistivity behavior versus field and we propose tunneling
of vortices as a plausible explanation for the observed result.
The standard tunneling probability for crossing a potential
barrier of height $V$ and width $d$ is proportional to
$exp[-k\sqrt{V}d]$ where $k$ is a constant. In the present case,
the barrier height is given by the activation energy of the vortex
$U_0$, and the barrier width by the distance between adjacent
pinning centers. We take an Anderson-Kim type activation energy
given by Yeshurun and Malozemoff $U_0\propto 1/H$ (see Eq. (1))
\cite{Yeshurun,Tinkham}, and assume a constant tunneling distance
between adjacent pinning centers in the weak link as this depends
on the specific material properties and not on the magnetic field.
This yields a flux flow tunneling current which is proportional to
the flux flow resistivity, that for any given temperature is given
by:
\begin{equation}
\rho_{t}=A exp[-D/\sqrt{H}]
\end{equation}
where $A$ and $D$ are constants. A fit of the data in Fig. 5 up to
2T using Eq. (2) is shown in this figure and its inset. A
reasonably good fit is now obtained for the low fields regime,
with a cross over to the Tinkham's model at higher fields. It
therefore seems that the Tinkham model is appropriate for films
with weak links or defects but not in the low fields regime. For
instance, we could fit the data of Kunchur \textit{et al.}
\cite{Kunchur} who used films with many defects and weak links
reasonably well, but again, not in the low field regime where
vortex tunneling apparently takes place (similarly to the
corresponding fit in Fig. 5). Also, the original Ambegaokar and
Halperin model \cite{Ambegaokar} which was used by Tinkham in his
model, was derived for a Josephson junction or a weak link. It is
thus not surprising that better fits are obtained
when weak links are involved.\\
\begin{figure}
\includegraphics[height=6.5cm,width=9cm]{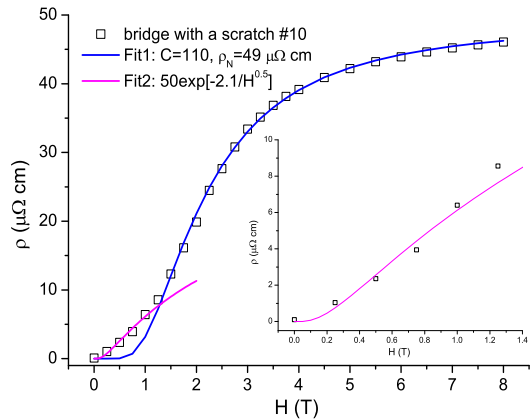}
\caption{\label{fig:epsart} (Color on-line) Main panel: Flux flow
resistivity of an antinode bridge with a weak link (a scratch) on
LAF552 versus applied magnetic field. The curves are fits to the
experimental data. Fit1 is obtained by using the Tinkham's model
of Eq. (1), and Fit2 is achieved by the tunneling model of Eq.
(2). Inset: zoom up on the low field regime with the tunneling
model fit.}
\end{figure}

\indent In conclusion, at temperatures close to $T_c$, low bias
and within the experimental error of our measurements, a clear
flux flow resistivity isotropy was observed in the present study
for the node and anti-node directions in the \textit{a-b} plane of
thin YBCO films. It is therefore demonstrated that the anisotropic
d-wave nature of the order parameter in YBCO does not induce any
measurable anisotropy in the flux pinning properties of the films.
We also show that the Ambegaokar and Halperin type model used by
Tinkham is more successful in films with weak links.\\

\indent We are grateful to Rudolf P. Huebener, Guy Deutscher, Assa
Auerbach and Lior Shkedy for useful discussions.  This research
was supported in part by the Israel Science Foundation (grant No.
1565/04), the Heinrich Hertz Minerva Center for HTSC, the Karl
Stoll Chair in advanced materials, and the Fund for the Promotion
of Research at the Technion.\\



\end{document}